\documentclass[12pt,letterpaper,english]{article}
\usepackage{amsmath}
\usepackage{amssymb}
\usepackage[small]{titlesec}

\makeatletter

\usepackage{amsfonts}

\usepackage{hyperref}

\numberwithin{equation}{section}
\newcommand{\hhref}[1]{\href{http://arxiv.org/abs/#1}{arXiv:#1}}

\begin{document}
\begin{titlepage}
\begin{flushright}
LPTENS-11/05

CALT 68-2819
\end{flushright}
\vskip 1.0cm
\begin{center}
{\Large \bf What Maxwell Theory in $d\neq4$ teaches us about scale
and conformal invariance}
\vskip 1.0cm
{\large Sheer El-Showk$^{a}$,\ Yu Nakayama$^{b}$, Slava Rychkov$^{c}$} \\[0.7cm]
{\it $^a $
Institut de Physique Th\'{e}orique, CEA Saclay,\\ and CNRS URA 2306, F-91191 Gif-sur-Yvette, France \\[5mm]
$^b$ California Institute of Technology, 452-48, Pasadena, CA 91125, USA
\\[5mm]
$^c$ Laboratoire de Physique Th\'{e}orique, \'{E}cole Normale Sup\'{e}rieure,\\
and Facult\'{e} de Physique, Universit\'{e} Pierre et Marie Curie,
France}
\end{center}
\vskip 1.0cm
\begin{abstract}
The free Maxwell theory in $d\ne4$ dimensions provides a physical example of a unitary, scale invariant theory which is NOT conformally invariant. The easiest way to see this is that the field strength operator $F_{\mu\nu}$ is neither a primary nor a descendant. We show how conformal multiplets can be completed, and conformality restored, by adding new local operators to the theory. In $d\geqslant 5$, this can only be done by sacrificing unitarity of the extended Hilbert space. We analyze the full symmetry structure of the extended theory, which turns out to be related to the $OSp(d,2|2)$ superalgebra.
\end{abstract}
\vskip 1cm \hspace{0.7cm} January 2011
\end{titlepage}

\newpage

\section{Introduction}

Conformal invariance plays a central role in quantum field theory
(QFT). It is often assumed to be an almost inevitable consequence
of scale invariance. For instance, in the world-sheet formulation
of string theory, the target space equations of motion can be
obtained by imposing the vanishing of the beta function of the underlying
two-dimensional world-sheet field theory. A vanishing beta function
only indicates scale invariance, and does not automatically imply
conformal invariance. Yet, it is the conformal invariance that is
necessary for the consistency of the world-sheet string theory. Can
we find a string background that satisfies the equations of motion
but is inconsistent just because of the lack of conformal invariance?
Another example arises in condensed matter physics. In order to classify the possible critical
phenomena, we use the conformal field theory (CFT) techniques. Is
it really the case that critical phenomena must always enjoy conformal
invariance while their definition only requires scale invariance?
Is there any deep physics behind the distinction between scale invariance
and conformal invariance?

That scale invariance should \emph{generically} imply conformal invariance
can be intuitively understood as follows. If a theory is scale invariant,
the trace of its energy-momentum tensor should be a total divergence \cite{Wess}\cite{Coleman:1970je}:
\begin{equation}
T_{\mu}^{\mu}=\partial_{\mu}k^{\mu}.\label{eq-scale}
\end{equation}
Here $k_{\mu}$ is the `virial current' (internal part of the scale
current). Since $T_{\mu\nu}$ has dimension $d$, $k_{\mu}$ must
have dimension exactly $d-1$ at the fixed point. However, barring
coincidences, only \emph{conserved} currents do not acquire anomalous
dimensions. Thus, generically, $k_{\mu}$ will be conserved, so that
(\ref{eq-scale}) implies \[
T_{\mu}^{\mu}=0,\]
 which is the condition for conformal invariance.

The word {}``generically'' in the above paragraph is important,
since scale but not conformally invariant theories do exist. One class
of examples is formed by non-unitary theories, such as the theory
of elasticity (i.e. the free vector field without gauge invariance)
\cite{Riva:2005gd}, or theories based on higher-derivative Lagrangians.
A second class includes theories which are unitary but do not have
an energy-momentum tensor operator, such as the linearized gravity, or a Gaussian
vector field with a non-conformally invariant two-point function \cite{Dorigoni:2009ra}.
A third class is based on non-compact models like the deformed Liouville
theory \cite{Iorio:1996ad}\cite{Ho:2008nr}, which typically break
unitarity as well.

It turns out that in $d=2$ spacetime dimensions, this classification
of counterexamples is complete: a fundamental theorem due to Zamolodchikov
and Polchinski \cite{Zamolodchikov:1986gt,Polchinski:1987dy}%
\footnote{See also \cite{Mack1} for an independent work on the theorem.%
} says that any scale invariant 2D QFT which is unitary, has a well-defined
 energy-momentum tensor, and has a discrete spectrum, is conformally invariant.

For $d\geqslant3,$ the situation remains unclear. Polchinski \cite{Polchinski:1987dy}
in 1987 undertook a detailed review of the pre-existing literature
and found no counterexamples. Systematic searches among theories having
candidates for a nonconserved virial current \cite{Polchinski:1987dy,Dorigoni:2009ra}
have not turned up any counterexamples either. Consequently, a general
conjecture seems to be that the Zamolodchikov-Polchinski theorem is true in $d\geqslant3$,
even though a proof is so far elusive.

In this note we explain that this conjecture is false, at least in
$d=3$ and in $d\geqslant5$. The counterexample is astonishingly
simple: it is the free Maxwell theory!
 The theory is unitary, has a well-defined energy-momentum tensor
and has a discrete spectrum, and it is legitimate to call it a physical
scale invariant but non-conformal field theory. Moreover, in $d\geqslant5$, we will
see a far richer structure of the theory by embedding it (in the BRST
sense) into a \textit{non-unitary} CFT, whose physical subsector coincides
with the unitary Maxwell theory.

The note is organized as follows. In Section 2 we briefly remind the
reader the axioms of conformal field theory, and demonstrate them on the $d=4$ Maxwell
theory, which is a bona fide CFT. In Sections 3,4,5 we discuss the
Maxwell theory in $d=3$ and $d\geqslant5$. These theories are not
conformal since $F_{\mu\nu}$, the lowest dimension gauge invariant
operator, is neither a primary nor a descendant. Interestingly, in
both cases conformality \emph{can} be saved but only at the price
of \emph{changing the theory} by extending the space of local
operators. In the extended Hilbert space $F_{\mu\nu}$ is a descendant.
The difference between $d=3$ and $d\geqslant5$ is that in the second
case the extended Hilbert space is, necessarily, not unitary. This
poses further conceptual puzzles, which we attempt to resolve. We
summarize and conclude in Section 6.

\textbf{Note added.} When our paper was being prepared for publication,
paper \cite{Jackiw:2011vz} appeared which also discusses scale and
conformal (non)invariance of the $d\ne4$ Maxwell theory, at the classical level.

\section{Maxwell theory in $d=4$ as Conformal Field Theory}

\label{sec:d4}In this paper we will consistently emphasize the arguments
based on correlation functions over the conventional Lagrangian methods.
On the most basic level, QFT is simply a set of correlators of local
operators. These correlators must satisfy well-known and natural axioms
(the Wightman axioms), such as Lorentz invariance, microcausality,
and unitarity, to which we add the existence of the energy-momentum tensor operator%
\footnote{The Wightman axioms assume the existence of the total energy-momentum
but not necessarily of its density.%
}. A Lagrangian is useful inasmuch as it allows one to \emph{compute}
the correlators; having a good Lagrangian also guarantees that the
correlators will satisfy the axioms. Once the correlators are computed,
however, and the theory thus defined, we can safely discard the Lagrangian
and never recall it.

Consider the free Maxwell theory in $d=4$ from this point of view.
The Lagrangian\[
\mathcal{L}=-\frac{1}{4}(F_{\mu\nu})^{2}\ \]
 defines the two-point function of the gauge potential $A_{\mu}$.
In the position space (we do not keep track of the overall field normalization)
\[
\langle A_{\mu}(x)A_{\nu}(0)\rangle=\frac{\eta_{\mu\nu}}{x^{2}}+\mbox{gauge terms }\qquad(d=4).\]
 This two-point function is gauge-variant and not physical, but it
serves to define the correlators of the gauge invariant field strength
$F_{\mu\nu}=\partial_{\mu}A_{\nu}-\partial_{\nu}A_{\mu}$. For example,
the two-point function of $F_{\mu\nu}$ comes out to be\begin{equation}
\langle F_{\mu\nu}(x)F_{\lambda\sigma}(0)\rangle=\frac{I_{\mu\lambda}I_{\nu\sigma}-\mu\leftrightarrow\nu}{(x^{2})^{2}},\quad I_{\mu\nu}\equiv\eta_{\mu\nu}-2\frac{x_{\mu}x_{\nu}}{x^{2}}\qquad(d=4).\label{eq:FF4}\end{equation}
 The latter set of gauge invariant correlators \emph{defines the theory}.
It is not yet the full theory because other local and gauge invariant
fields can be constructed by taking the OPE of $F_{\mu\nu}$ with
itself. Since we are dealing with a free theory, these fields are
nothing but normal ordered powers of $F_{\mu\nu}$. Among them, two notable
fields are the scalar\[
\Phi=\,:\negmedspace(F_{\mu\nu})^{2}\negmedspace:\]
 and the symmetric tensor \begin{align}
T_{\mu\nu}=\,:\negmedspace F_{\mu\rho}F_{\ \nu}^{\rho}-\frac{\eta_{\mu\nu}}{4}(F_{\rho\sigma})^{2}\!:\ .\label{emmax}\end{align}
 The latter field is in fact the energy-momentum tensor of the theory. In textbook
treatments, its expression is obtained from the Lagrangian, e.g.\ by
varying the action with respect to the external metric. Notice, however,
that this is not strictly speaking necessary. Instead, one could have
simply checked that the correlators of $T_{\mu\nu}$ defined by \eqref{emmax}
are conserved and satisfy the Ward identities.

The $d=4$ Maxwell theory is obviously scale invariant (all correlators
scale with distance). As is well known, it is also conformally invariant.
The fastest way to see this is to observe that the energy-momentum tensor (\ref{emmax})
is traceless. What if we did not know the explicit expression of the
 energy-momentum tensor? How would we then decide if the theory is conformal
or not?

The answer is: look at the correlators and see if they are consistent
with conformal symmetry. In a CFT, local fields are classified into
primaries and their derivatives (descendants). Correlators of primaries
are strongly constrained by conformal invariance. For example, the
correlator of three scalar primaries $\phi_{i}$ of equal dimensions
$\Delta_{i}=\Delta$ must take the form
\[
\langle\phi_{i}(x_{1})\,\phi_{j}(x_{2})\,\phi_{k}(x_{3})\rangle=
\frac{c_{ijk}}{[(x_{1}-x_{2})^{2}(x_{2}-x_{3})^{2}(x_{3}-x_{1})^{2}]^{\Delta/2}}\ .\]
 Following this logic, let us compute the three-point function of the $\Phi=\,:\negmedspace(F_{\mu\nu})^{2}\negmedspace:$. We find that it vanishes:
\[
\langle\Phi(x)\Phi(y)\Phi(z)\rangle=0\qquad(d=4).\]
 This equation is consistent with $\Phi$ being a scalar primary,
with an (accidentally) vanishing three-point function coefficient $c_{\Phi\Phi\Phi}$.\footnote{This fact was also noticed in \cite{DO} by studying the conformal block decomposition of the four-point function of $\Phi$.}

A further constraint is that the two-point functions of tensor primaries
have to be constructed out of the $I_{\mu\nu}$ tensor (see e.g.\ \cite{Osborn}). Eq. (\ref{eq:FF4})
is thus consistent with $F_{\mu\nu}$ being an antisymmetric tensor
primary.

It is good news for the conformal invariance of the theory that both
$\Phi$ and $F_{\mu\nu}$ can be interpreted as primaries. Had we
found a contradiction at this point, the only remaining possibility
(consistent with conformality of the theory) would be to try to interpret
them as descendants. However, there are no candidate gauge invariant
local fields in the theory of which $\Phi$ and $F_{\mu\nu}$ could
be descendants.

Let us finally discuss unitarity in some detail. Unitarity is an important
part of the Wightman axioms, and means positivity of the spectral
density for Minkowski-space non time-ordered correlators.%
\footnote{In the Euclidean signature unitarity is replaced by the (Osterwalder-Schraeder)
reflection positivity.%
} Applied to two-point functions $\langle0|O^{\dagger}(x)O(y)|0\rangle$,
unitarity implies lower bounds on the scaling dimensions of local
operators. For example, scalars have dimensions $\Delta\ge1$. In
conformal theories, unitarity bounds for each primary type (vector,
symmetric traceless and antisymmetric two-tensors, etc.) can be obtained \cite{mack}.
The dimension of an antisymmetric tensor primary is always $\Delta\ge2,$
saturated by $F_{\mu\nu}$. The vectors have dimensions $\Delta\ge3$.
There is no contradiction that the Maxwell gauge potential violates
this bound, since it does not even belong to the Hilbert space of
the theory (let alone being a primary).

\section{Maxwell theory in $d\ne4$. Conformality lost.}

\label{sec:conf.lost}Let us now consider the Maxwell theory in $d\geqslant3, d\neq4$.\footnote{The $d=2$ Maxwell theory has no propagating degrees of freedom.%
} 
By the Maxwell theory we mean the same as in the previous section, i.e.
the set of correlation functions of the gauge invariant fields, such
as $F_{\mu\nu}$ and its operator products. The gauge potential is
not a physical field of the theory: its gauge dependent two-point
function \[
\langle A_{\mu}(x)A_{\nu}(0)\rangle=\frac{\eta_{\mu\nu}}{(x^{2})^{(d-2)/2}}+\mbox{gauge terms}\]
 is only used to define the correlators of $F_{\mu\nu}$.

It is obvious that the $d\ne4$ Maxwell is scale invariant; it is
also unitary, just like in $d=4$. Is it conformally invariant? If
we are to follow the textbook approach, we have to compute the trace
of the energy-momentum tensor (\ref{emmax}). We find \begin{align}
T_{\ \mu}^{\mu}=\frac{4-d}{4}(F_{\mu\nu})^{2}, \end{align}
nonzero in $d\ne4$. However, by itself this does not
automatically imply the absence of conformal invariance. For example,
the free scalar energy-momentum tensor \[
T_{\mu\nu}^{\phi}=\partial_{\mu}\phi\partial_{\nu}\phi-\frac{1}{2}\eta_{\mu\nu}(\partial\phi)^{2}\]
 suffers from the same problem if $d\ne2$, yet it is known that it
can be `improved', so that the free scalar theory is in fact conformally
invariant in any dimension. This is because the virial current happens to be
a total divergence (see below).

In a short while, we will discuss the impossibility of improving
(\ref{emmax}), unlike in the scalar case. But first, let us look at the
two-point function of $F_{\mu\nu}$ and the three-point function of $\Phi=\,:\negmedspace(F_{\mu\nu})^{2}\negmedspace:$.
For a general $d$ they turn out to have the form:
\begin{equation}
\left\langle F_{\mu\nu}(x)F_{\lambda\sigma}(0)\right\rangle =\frac{2d-4}{(x^{2})^{d/2}}\left[\left(\eta_{\mu\lambda}-\frac{d}{2}\frac{x_{\mu}x_{\lambda}}{x^{2}}\right)\left(\eta_{\nu\sigma}-\frac{d}{2}\frac{x_{\nu}x_{\sigma}}{x^{2}}\right)-\mu\leftrightarrow\nu\right],\label{eq:FF-1}
\end{equation}
 \begin{align}
\left\langle \Phi(x_{1})\Phi(x_{2})\Phi(x_{3})\right\rangle  & =\frac{-8(d-2)^{3}(d-4)d}{(x_{12}^{2})^{d/2}(x_{13}^{2})^{d/2}(x_{23}^{2})^{d/2}}\times\nonumber \\
 & \left[2+d^{2}\frac{(x_{12}.x_{13})(x_{12}.x_{23})(x_{13}.x_{23})}{x_{12}^{2}x_{13}^{2}x_{23}^{2}}-d\frac{(x_{12}.x_{23})x_{13}^{2}+2\text{ perms}}{x_{12}^{2}x_{13}^{2}x_{23}^{2}}\right]\ \label{F2-1}\end{align}
 (here $x_{ij}\equiv x_{i}-x_{j}$).

For $d\ne4,$ these equations imply that neither $F_{\mu\nu}$ nor
$\Phi$ can be conformal primaries. If the $d\ne4$ Maxwell is to be conformal, these fields have
to be descendants of some other local gauge invariant fields. But
of which ones? There are no other gauge invariant fields of sufficiently
low dimension which could serve this purpose. In fact, $F_{\mu\nu}$
is the lowest-dimensional gauge invariant field, so it cannot be anyone
else's descendant. The $\Phi$ and $T_{\mu\nu}$ are the next-to-lowest
dimensional fields after $F_{\mu\nu}$. Thus $\Phi$ could only be
a descendant of $F_{\mu\nu}$. But $F_{\mu\nu}$ has no scalar descendants,
since $\partial^{\mu}F_{\mu\nu}=0$.

We are led to conclude that the $d\ne4$ Maxwell is not conformally invariant.

For completeness, let us discuss how the same conclusion could be
reached using the textbook approach, by showing that the energy-momentum tensor
cannot be improved. To begin with, note that the energy-momentum tensor trace
can be represented as a total derivative
\[
T_{\mu}^{\mu}=\frac{4-d}{8}\partial_{\mu}k^{\mu},\qquad k^{\mu}=A_{\rho}F^{\rho\mu}\ .\]
 This had to happen since the theory is scale invariant. Notice that it
is somewhat peculiar that the virial current $k_{\mu}$ is not gauge invariant%
\footnote{The assumption that $k_{\mu}$ must be a physical, gauge invariant
(or BRST invariant), operator was implicit in \cite{Polchinski:1987dy}.
The $d\ne4$ Maxwell theory apparently violates this assumption.%
}. As a consequence, the conserved dilatation current $J_{\mu}^{D}=x^{\nu}T_{\nu\mu}-k_{\mu}$
is not gauge invariant. However, one can check that the dilatation
charge $D=\int d^{3}xJ_{0}^{D}$ is gauge invariant. Using the equations
of motion, \begin{align}
\delta_{\Lambda}D=\int d^{3}x\,\partial^{i}\Lambda F_{i0}=-\int d^{3}x\,\Lambda\partial^{i}F_{i0}=0\ .\end{align}

The condition for the improvement of the energy-momentum tensor (so
that it can be traceless) is that the $k_{\mu}$ be a total derivative
\cite{Coleman:1970je,Polchinski:1987dy}: \begin{align}
k^{\mu}=\partial^{\nu}L_{\nu\mu}\ ,\end{align}
 where $L_{\mu\nu}$ must be a dimension $d-2$ symmetric tensor constructed
out of the local fields of the theory (for $d>2$). On dimensional grounds
the only possibility is  \[
L_{\mu\nu}=a_{1}A_{\mu}A_{\nu}+a_{2}\eta_{\mu\nu}(A_{\rho})^{2},\]
 but it is easy to check that this cannot generate the above $k_{\mu}$
no matter how we choose $a_{1}$ and $a_{2}$.

\section{Recovering conformality in $d=3$}

As we have seen in the previous section, the $d\ne4$ Maxwell theory
is scale invariant but not conformal, since there are gauge invariant
local fields which are neither primaries nor descendants.

It turns out that a partial fix is possible. Namely, we can try to
add new local fields to the theory. The new extended theory will contain
all the correlation functions of the original Maxwell theory and the
correlators of the new fields. The new fields will be primaries, and
the old fields like $F_{\mu\nu}$ or $\Phi$ will now be descendants
of the new fields. Thus the extended theory will be conformal.
\footnote{Another way to turn $d\neq 4$ Maxwell theory into a classically
conformal theory would be to couple it to a conformal compensator scalar field,
i.e.\ by considering actions of the form $F_{\mu\nu}^2
\phi^\alpha+(\partial\phi)^2$ for an appropriate $\alpha$. We are grateful to
A.~Tseytlin for this remark. Notice however that at the quantum level, such an
action could describe only an effective field theory in a phase of spontaneously
broken conformal invariance (since one would have to give $\phi$ a vev). There
would be severe difficulties in defining such a theory in the UV. In this paper
we deal with much simpler theories, but with the virtue that they make sense
quantum-mechanically at all energies.}

This procedure turns out to work differently in $d=3$ and $d\ge5$;
in this section we focus on $d=3.$

Let us add to the theory a free scalar field $B$. We will postulate
that $F_{\mu\nu}$ is a descendant of $B,$ according to the formula\begin{equation}
F_{\mu\nu}=\epsilon_{\mu\nu\rho}\partial^{\rho}B\ .\label{eq:FB}\end{equation}
 Physically, $B$ is the non-local magnetic dual of the gauge potential.

It is not difficult to check that, in $d=3,$ prescription (\ref{eq:FB})
gives the $F_{\mu\nu}$ two-point function that is identical to Eq.
(\ref{eq:FF-1}). Furthermore, the extended theory will also contain
new composite fields constructed out of $B$. Their existence is essential
to complete all conformal multiplets. For example, $\Phi$ will now
be a descendant of $:\! B^{2}\!:$.

The outlined construction can be rephrased as follows: \emph{In $d=3$,
free scalar theory contains a subsector which is isomorphic to the
Maxwell theory}. By a subsector we mean here a set of correlators
closed with respect to the OPE.

In other words, we have saved conformality by extending the $d=3$
Maxwell into the free scalar theory, which is conformal for any $d$.
It has to be stressed that by passing from Maxwell to free scalar
we really changed the theory (changed the set of local operators).
The extended theory is unitary, so we can successfully embed the non-conformal
Maxwell theory in $d=3$ into a unitary CFT. It
would be interesting to understand if there is another reason (except
for the need to recover conformal invariance) which would justify
such an extension.

One idea would be to see if modular invariance of the torus partition
function requires this. This would then be analogous to the Ising
model in $d=2$, where the OPE closes in the $\epsilon$ sector, yet
modular invariance makes the presence of the $\sigma$ field mandatory
\cite{Cardy1986}.

In $d=3$, another example which comes to mind is the M2-brane gauge
theory (i.e. the ABJM model \cite{Aharony:2008ug}). At levels $k=1,2$,
$\mathcal{N}=8$ supersymmetry is not manifest in the original variables
appearing in the action of this theory. We have to introduce magnetic
monopole operators (essentially the same field $B$) to construct
$\mathcal{N}=8$ SUSY multiplets (see e.g. \cite{Gustavsson:2009pm}).
In this example, the manifestation of the maximal supersymmetry
required the existence of the magnetic dual operators.

\section{Recovering conformality in $d\ge5$}

\subsection{No unitary extension}

We will now discuss whether the $d\ge5$ Maxwell can also be extended
into a conformal theory, similarly to $d=3$. To answer this question,
we have to try to add a field that satisfies the following property:
it has the two-point function of a primary field, and we can construct
$F_{\mu\nu}$ as its descendant. However, this runs into the following
difficulty. Taking into account that all bosonic primaries are symmetric
traceless or antisymmetric tensors, there are only two possible descendant relations:
\begin{align*}
F_{\mu\nu} & =\partial_{\mu}Y_{\nu}-\partial_{\nu}Y_{\mu}\ ,\\
F_{\mu\nu} & =\epsilon_{\mu\nu\lambda\ldots}\partial^{\lambda}Z^{\ldots}\ ,\end{align*}
 where $Y$ is a hypothetical primary vector (not to be confused with
the vector potential $A_{\mu}$), while $Z$ would be a totally
antisymmetric tensor of rank $d-3$. To
be consistent with the scaling dimension of $F_{\mu\nu}$, $\Delta_{F}=d/2$,
these fields would have to have dimension $d/2-1$. The trouble is, such low dimensions
are inconsistent with unitarity. Unitarity bounds for higher dimensional
conformal fields theories were derived in \cite{Metsaev95},\cite{Minwalla:1997ka}\footnote{Metsaev \cite{Metsaev95} used 
a field-theoretic AdS realization which gives results equivalent to a purely group-theoretic derivation given later by Minwalla \cite{Minwalla:1997ka}.};
in the cases of interest for us they read:
\begin{align}
\Delta & \ge d-1\qquad\mbox{ (primary vector)}\label{eq:minw1}\\
 \Delta & \ge \max(d-k,k)\qquad \mbox{(primary rank \ensuremath{k}\ antisymmetric tensor)}\label{eq:minw2}
 \end{align}
 We are thus led to the conclusion that it is impossible to extend the $d\ge5$ Maxwell into a \emph{unitary}
conformal theory.

\subsection{A non-unitary extension}

Despite the difficulty we have mentioned, it turns out that an extension
into a \emph{non-unitary} conformal theory is possible. Namely, let
us add to the theory a local field $Y_{\mu}$ having a two-point function
of a dimension $d/2-1$ vector primary:
\[
\langle Y_{\mu}(x)Y_{\nu}(0)\rangle=\frac{I_{\mu\nu}}{(x^{2})^{d/2-1}}\ .\]
 Assume that this field is Gaussian, i.e. all higher order correlators
are computed via Wick's theorem. This, then, is a conformal field
theory, which is non-unitary, since the unitarity bound (\ref{eq:minw1})
is violated (see Appendix \ref{sec:u-bound} for an elementary derivation
of this bound).

Assume further that $F_{\mu\nu}$ is a descendant of $Y_{\mu}$ via\begin{equation}
F_{\mu\nu}=\partial_{\mu}Y_{\nu}-\partial_{\nu}Y_{\mu}\ .\label{eq:FY}\end{equation}
 It is not difficult to check that this ansatz reproduces the two-point
function of $F_{\mu\nu}$ given by (\ref{eq:FF-1}), up to a normalization
factor of $(d-4)/(d-2)$. This is all we need to demonstrate the extension.

This result suggests several interesting questions, to be discussed
below.

\subsection{The origin of $Y_{\mu}$}

In Sections \ref{sec:d4},\ref{sec:conf.lost} we have not specified
the gauge in which we compute the gauge field propagator (since, of
course, all gauges give the same $F_{\mu\nu}$ two-point functions).
Let us now focus on the $\xi$-gauge,\begin{equation}
\mathcal{L}_{\xi}=-\frac{1}{4}(F_{\mu\nu})^{2}-\frac{1}{2\xi}(\partial_{\mu}A^{\mu})^{2}\ ,\ \label{eq:Lxi}\end{equation}
 so that the momentum space propagator takes the well-known form\[
\langle A_{\mu}(-p)A_{\nu}(p)\rangle=\frac{\eta_{\mu\nu}}{p^{2}}\left(1-(1-\xi)\frac{p_{\mu}p_{\nu}}{p^2}\right)
\ ,
\]
 from which the coordinate-space propagator is found to be\[
\langle A_{\mu}(x)A_{\nu}(0)\rangle=\frac{1}{(x^{2})^{(d-1)/2}}\left(\eta_{\mu\nu}+(d-2)\frac{1-\xi}{1+\xi}\frac{x_{\mu}x_{\nu}}{x^{2}}\right)\]
 (up to an overall constant factor).

Now notice a curious thing. For a particular choice of the gauge parameter\[
\xi_{*}=\frac{d}{d-4}\]
 the two-point function of $A_{\mu}$ becomes proportional to the $I_{\mu\nu}$
tensor, and thus takes the form consistent with $A_{\mu}$ being a
conformal primary. For this value of $\xi$, the gauge-fixed theory
(\ref{eq:Lxi}) is conformal. (In Appendix \ref{sec:EMxi} we also
show that its energy-momentum tensor can be improved to become traceless.)

This, then, is the underlying reason which made it possible to extend 
the Maxwell theory into a CFT in which the field strength is a descendant
of a primary vector field. The primary vector $Y_{\mu}$ is nothing
but the vector potential $A_{\mu}$ in a particular $\xi$-gauge.
This of course guarantees that the identification (\ref{eq:FY}) will
produce the correct $F_{\mu\nu}$ two-point function.

\subsection{BRST-like interpretation}

From now on we will consider the theory (\ref{eq:Lxi}) with $\xi=\xi_{*}$
and will rename $A_{\mu}$ to $Y_{\mu}$, in order to avoid any possible confusion.
As discussed above, this is a CFT, however of an unusual sort. While
the full theory is non-unitary, there is a subsector of the theory
(the Maxwell theory) which is unitary. The conformal multiplet of which $Y_{\mu}$
is the primary contains both non-unitary ($Y_{\mu}$ itself) and unitary
$(F_{\mu\nu}$ and its derivatives) fields. This is clearly not a
bona fide unitary CFT, where all local fields must belong to the unitary
Hilbert space.

One would like to come up with a procedure to distinguish the unitary
sector from non-unitary sectors of the theory. As usual in gauge-fixed
theories, BRST is the way to implement the distinction. Thus, let
us add to the theory the ghost sector\begin{equation}
\mathcal{L}_{gh}=-\frac{1}{2}\epsilon_{ab}c_{a}\partial^{2}c_{b}\ .\label{eq:gh}\end{equation}
 On the one hand, the new local fields $c_{1,2}$, anticommuting scalars,
also violate unitarity. On the other hand, the full theory now has
a BRST-like symmetry with two fermionic generators (`BRST' and
`anti-BRST')\begin{equation}
[Q_{a},Y_{\mu}]=-i\partial_{\mu}c_{a},\qquad\{Q_{a},c_{b}\}=-i\xi_{*}^{-1}\epsilon_{ab}\partial_{\mu}Y_{\mu}.\label{eq:BRST}\end{equation}
 The unitary subsector of the theory can now be picked out by saying
that it consists of all $Q$-invariant fields.%
\footnote{One of the $Q_{a}$ or both of them would equally work.%
}

There is one curious thing about the above construction which merits
further attention: the BRST transformations do not commute with the
conformal algebra. This is already obvious from the fact
that a BRST-variant primary field $Y_{\mu}$ has a BRST-invariant
descendant $F_{\mu\nu}$. This looks pretty unusual: for example in
string-theoretic 2D CFTs the BRST operator always commutes with the conformal
algebra. This also provokes us to try to identify the full symmetry
(super)algebra of the theory, containing both the BRST and conformal generators.

\subsection{Extended conformal algebra}

Conformal algebra commutation relations and the generator action on
the primary fields are summarized in Appendix \ref{sec:algebra}.
In particular, the action on the ghosts $c_{a}$, which are scalars, is
given by Eq. (\ref{eq:gen.action}) with $\Delta=(d-2)/2$ and $\Sigma\equiv0$.
The action on the primary vector $Y_{\mu}$ is given by the same equations
but now $\Sigma$ is nontrivial: $(\Sigma_{\mu\nu}Y)_{\rho}=\eta_{\mu\rho}Y_{\nu}-\eta_{\nu\rho}Y_{\mu}$.

In order to understand the full symmetry algebra, we have to include
Eq. (\ref{eq:BRST}), which defines the action of fermionic generators
$Q_{a}$.

As already mentioned above, $Q$'s cannot commute with the conformal
algebra: the relations\footnote{In the section $\sim$ means equality up to an irrelevant $c$-number factor.} 
\[
[Q_{a},F_{\mu\nu}]=0,\qquad[Q_{a},Y_{\mu}]\ne0,\qquad Y_{\mu}\sim[K_{\nu},F_{\mu\nu}]\]
 can only be consistent if 
 \begin{equation}
Q_{a\mu}\equiv i[Q_{a},K_{\mu}]\ne0.\label{eq:ferm.rot.}
\end{equation}
 An explicit computation shows that, indeed, these new `fermionic rotation'
generators $Q_{a\mu}$ act nontrivially on the primary fields:\begin{align*}
\{Q_{a\mu},c_{b}(x)\} & =-i\xi_{*}^{-1}\epsilon_{ab}[d\,\eta_{\mu\nu}-2x_{\mu}\partial_{\nu}]Y_{\nu}(x)\ ,\\
{}[Q_{a\mu},Y_{\nu}(x)] & =i[(d-2)\eta_{\mu\nu}+2x_{\mu}\partial_{\nu}]c_{a}(x)\,.
\end{align*}
What is, then, the full symmetry (super)algebra of the theory we are
dealing with?

A natural conjecture could be that this is the orthosymplectic superalgebra
$OSp(d,2|2)$. Indeed, this superalgebra has the bosonic subgroup
$SO(d,2)\times Sp(2)$, and is the smallest (simple) superalgebra
that contains this symmetry \cite{Kac}. The first factor could then
be identified with the conformal algebra, while the second factor
acts on the ghosts which form the fundamental multiplet of $Sp(2)$,
so that the ghost action (\ref{eq:gh}) is invariant. The full (anti)commutation
relations of $OSp(d,2|2)$ are \cite{Barducci:1987hu}
\begin{align*}
\lbrack J_{AB},J_{CD}\} & =i\bigl(\eta_{CB}J_{AD}-(-1)^{[A][B]}\eta_{CA}J_{BD}+(-1)^{[A]([B]+[C])}\eta_{DA}J_{BC}-(-1)^{[B][C]}\eta_{DB}J_{AC}\bigr),\\[5pt]
 \eta_{AB}&=(\eta_{\alpha\beta},\varepsilon_{ab}),  \ \alpha,\beta=0,\ldots,d+1,\ a,b=1,2;\quad\text{gradings }[\alpha]  =0,[a]=1.\end{align*}
 Here $J_{[\alpha\beta]}$ are the $SO(d,2)$ generators (see Appendix
\ref{sec:algebra}), $J_{ab}(ab=11,12,22)$ are the $Sp(2)$ generators.
The fermionic generators are $J_{a\pm}$ ($x^{\pm}=x^{d+1}\pm x^{d})$
and $J_{a\mu}$; they satisfy the commutation relations
\begin{align}
[J_{a+},J_{\mu-}] & =-\frac{i}{2}J_{a\mu}\ ,\label{eq:comm1}\\
{}[J_{a\mu},J_{\nu-}] & =i\eta_{\mu\nu}J_{a-}\ ,\label{eq:comm2}\\
\{J_{a\alpha},J_{b\beta}\} & =i(\eta_{\alpha\beta}J_{ab}-\varepsilon_{ab}J_{\alpha\beta})\ .\label{eq:comm3}
\end{align}
 Recall that $K_{\mu}\sim J_{\mu-}$, $P_{\mu}\sim J_{\mu+}$. Let us try to identify $Q_{a}\sim J_{a+}$
so that the BRST operator becomes a sort of fermionic
$P_{\mu}$. This is consistent with $\{Q_{a},Q_{b}\}=0$. Then Eq.\ (\ref{eq:comm1}) identifies $J_{a\mu}$ with the fermionic rotation
$Q_{a\mu}$ as defined above.

However, unfortunately, this $OSp(d,2|2)$ conjecture does not hold
up since the algebra with the above identifications does not close.
To see this, consider the partial case of Eq.\ (\ref{eq:comm3}),\[
\{Q_{a},Q_{b\mu}\}\sim i\epsilon_{ab}P_{\mu}\qquad(?)\]
 On the other hand, an explicit computation using the known action
of $Q_{a}$ and $Q_{b\mu}$ on the fields shows: 
\begin{align}
\{Q_{a},Q_{b\mu}\}=i\epsilon_{ab}(d-4)(P_{\mu}+P{}_{\mu}'),\ \label{bad}
\end{align}
 where $P_{\mu}'$ is a new bosonic generator which acts only
on $Y_{\mu}$ according to\[
[P'_{\mu},Y_{\nu}(x)]=-i[\partial_{\nu}Y_{\mu}-\partial_{\mu}Y_{\nu}+\xi_{*}^{-1}\eta_{\mu\nu}(\partial Y)]\ .\]
 These new generators do not commute: \begin{align}
P'_{[\mu\nu]}\equiv i [P'_{\mu},P'_{\nu}]\ne0,\end{align}
 where $P'_{[\mu\nu]}$ generate higher spin symmetries of the theory
of the form \begin{align}
\delta Y_{\mu}=a_{[\mu\nu]}\partial_{\nu}(\partial^{\rho}Y_{\rho})+\xi_{*}a_{[\rho\sigma]}\partial_{\sigma}\partial^{\mu}Y^{\rho}\ .\end{align}
 By iterating this procedure $[P'_{\mu_{1}},[P'_{\mu_{2}},[P'_{\mu_{3}},\cdots]]]$,
one can construct an infinite number of antisymmetric tensor conserved
currents. One can also construct their fermionic counterparts by commuting
with the (anti-)BRST charges $Q_{a}$.

Of course, since the theory we are dealing with is free, we should
not be surprised by the presence of infinite-dimensional higher spin
symmetry algebras. It was not obvious that these currents must be
generated by repeated commutators of $Q_{a}$ and $K_{\mu}$ (because
$OSp(d,2|2)$ could be a minimal closure of the algebra), yet the
above discussion shows that the generation of the higher spin symmetry
does happen in our theory.

Notice that for the above conclusion it was important that we included
both BRST and anti-BRST generators into the symmetry algebra. Were
we to artificially leave out one of them, the closure would be a finite-dimensional
super-algebra which is a subalgebra of $OSp(d,2|2)$. This subalgebra
would include the conformal generators $J_{\alpha\beta}$, the BRST
charge $Q_{B}=J_{1-}$, the fermionic vector charges $J_{1\mu}$ and
the fermionic scalar charge $J_{1+}$ (here $1$ denotes a component
of the $Sp(2)$ index $a$). The point is that the $OSp(d,2|2)$ has
a grading with respect to the ghost number generator $J_{12}$ within
$Sp(2)$. We can thus construct a subalgebra by restricting to the
non-negative ghost number sector. In particular, the problematic anti-commutation
\eqref{bad} is avoided, and $P'$ would not appear in the algebra.
This subalgebra is not simple (and thus does not appear in the Kac
classification), because it contains a non-trivial ideal consisting
of the strictly positive ghost number generators.

\section{Summary, discussion and lessons}

In this paper, we started by showing the following three technical
results: 
\begin{itemize}
\item \emph{The Maxwell theory in $d\ne4$ is an example of a unitary, scale invariant
theory which is not conformally invariant.} We demonstrated this
in two ways: a) by showing that the physical, gauge invariant fields
$F_{\mu\nu}$ and $:\!(F_{\mu\nu})^{2}\!:$ are neither primaries
nor descendants; b) in the textbook way - by showing that the energy-momentum
tensor cannot be improved to be traceless. 
\item \emph{The $d=3$ Maxwell theory can be extended to a unitary CFT}
by introducing a new local field: a primary scalar $B$. In the extended
theory, $F_{\mu\nu}$ is a descendant of $B$. 
\item \emph{The $d\ge5$ Maxwell theory }\textbf{\emph{cannot}}\emph{ be
extended to a unitary CFT}, since this would contradict unitarity
bounds. \emph{However, it can be extended to a }\textbf{\emph{non-unitary}}\emph{
CFT}, by introducing a primary vector $Y_{\mu}$ of which $F_{\mu\nu}$
is a descendant. 
\end{itemize}
This group of results is interesting for two reasons. First, it provides
a clearcut counterexample to the often assumed conjecture that any
unitary and scale invariant theory is conformal. It is pretty amazing
that this counterexample escaped the attention of the recent literature
on the subject.%
\footnote{We have noticed that Birrell and Davis \cite{Birrell:1982ix},
Section 3.8, do mention in passing that the Maxwell theory is conformally
invariant only in $d=4$. %
Another precursor result is a classification, for any $d$, of all primary operators $\phi$ which can 
consistently satisfy the free field equation $\partial^2\phi=0$ \cite{Siegel},\cite{Metsaev},\cite{Minwalla:1997ka}. 
An antisymmetric two-tensor is not in the list for $d\ne4$.
} Second, our results point a way to generalize the conjecture so that
it has a chance to remain true: Could it be that in fact any unitary
and scale invariant theory \emph{can be made} conformal by extending
the set of local fields?

At present we do not have any evidence in favor of this \emph{generalized
conjecture}  except for the fact that it does not contradict
our examples. A most puzzling thing is that non-unitary conformal
extensions need to be allowed, as the $d\ge5$ Maxwell theory case
shows. We thus proceeded to take a closer look at this case. Our findings
here can be summarized as follows: 
\begin{itemize}
\item We showed how to separate the unitary part of the Hilbert space by
using the usual BRST language. In fact, the non-unitary field $Y_{\mu}$
can be interpreted as the vector potential in a particular $\xi$-gauge
in which its propagator is conformally invariant. 
\item The (anti-)BRST operators do not commute with the conformal algebra. 
\item An attempt to close the algebra starting from BRST, anti-BRST and conformal
generators leads to an infinite-dimensional superalgebra which includes
higher-spin symmetry generators (present in the Maxwell theory since
it is free). On the other hand, starting from just the BRST and conformal generators
(i.e.\ omitting anti-BRST) we can generate a finite-dimensional algebra,
which is a subalgebra of the orthosymplectic superalgebra $OSp(d,2|2)$. 
\end{itemize}
Based on this study, we can speculate that if other, and interacting,
examples of theories following the above \emph{generalized conjecture}
with a non-unitary CFT extension do exist, they should also perhaps
realize $OSp(d,2|2)$ or its nonnegative ghost-number subalgebra in
their Hilbert space.

Unfortunately, it seems quite difficult to search for such counterexamples,
since it is impossible to marginally deform the $d\ge5$ Maxwell theory
by coupling it to matter. Indeed, the dimension of $A_{\mu}$ is different
from that of $\partial_{\mu}$ so that the covariant derivative $\partial_{\mu}+ieA_{\mu}$
with a dimensionless coupling $e$ cannot be introduced while preserving
scale invariance. The higher derivative couplings like the Pauli interactions
are always irrelevant, so it is impossible to classically deform the
$d\ge5$ Maxwell theory while preserving even the scale invariance.
In this sense, the $d\ge5$ Maxwell theory looks like an isolated
IR fixed point.

This would not exclude the possibility that the UV fixed point may
be asymptotically safe, and we might expect a scale invariant but non-conformal
field theory with a strongly coupled fixed point. Such a theory
may or may not possess the hidden non-unitarity realized conformal
invariance of our free Maxwell theory. In such a case, the anti-BRST
transformation and its commutation relation with the conformal charge
must be modified compared with the free Maxwell theory because we
would not expect an infinite number of conserved charges in interacting
field theories.

Another obstruction to constructing such theories comes from their dual holographic
descriptions (when they are expected to exist). It was proven that as long as
the effective supergravity description is valid with matter satisfing the strict
null energy condition, the solutions of the equation of motion (even with higher
derivative corrections) automatically possess the AdS isometry if we impose the
Poincar\'e isometry and the scaling isometry \cite{Nakayama:2009fe}. One could
argue that a free theory is not expected to have a dual that is in any sense
accurately described by supergravity, as the dual of a free theory is
necessarily very strongly coupled.  Nevertheless these holographic arguments
do suggest that it may be difficult to construct intereacting examples of
scale invariant but non-conformal theories.

Finally, we should stress that our examples are strictly in $d\geqslant3,\ d\ne4$.
The $d=4$ case of the Zamolodchikov-Polchinski theorem remains open; it could still be
that in this dimension, like in $d=2$, scale invariance $+$ unitarity
$\Longrightarrow$ conformal invariance.

\section*{Acknowledgements}

S.~R. is grateful to Jan Troost, Manuela Kulaxizi and Andrei Parnachev
for useful discussions. Y.~N. thanks John Cardy for
the encouraging and stimulating correspondence on the subject.
S.~E. would like to thank the CEA Saclay for hospitality during the completion of part of this work.  
The research of S.E. is partially supported by the Netherlands Organisation for Scientific Research (NWO) under a Rubicon grant.
The work of S.~R. was supported in part by
the European Programme {}``Unification in the LHC Era"{}{},
contract PITN-GA-2009-237920 (UNILHC). The work of Y.~N. is supported by Sherman Fairchild Fellowship at California Institute of Technology.

\appendix

\section{Conformal algebra}

\label{sec:algebra}Recall the conformal algebra%
\footnote{To avoid any possible confusion, let us note that the special conformal generator $K_{\mu}$
has nothing to do with the virial current $k_{\mu}$.%
} \begin{align}
[M_{\mu\nu},M_{\rho\sigma}] & =-i(\eta_{\mu\rho}M_{\nu\sigma}\pm\mbox{perms})\nonumber \\
{}[M_{\mu\nu},P_{\rho}] & =i(\eta_{\nu\rho}P_{\mu}-\eta_{\mu\rho}P_{\nu})\nonumber \\
{}[D,P_{\mu}] & =-iP_{\mu}\label{eq:CA}\\
{}[D,K_{\mu}] & =+iK_{\mu}\nonumber \\
{}[P_{\mu},K_{\nu}] & =2i(\eta_{\mu\nu}D-M_{\mu\nu})\ .\nonumber \end{align}
 The generators act on the primary fields (not necessarily scalars)
by \begin{align}
[P_{\mu},\phi(x)] & =-i\partial_{\mu}\phi(x)\nonumber \\
{}[D,\phi(x) & ]=-i(\Delta+x^{\mu}\partial_{\mu})\phi(x)\label{eq:gen.action}\\
{}[M_{\mu\nu},\phi(x)] & =\{\Sigma_{\mu\nu}-i(x_{\mu}\partial_{\nu}-x_{\nu}\partial_{\mu})\}\phi(x)\nonumber \\
{}[K_{\mu},\phi(x)] & =(-i2x_{\mu}\Delta-2x^{\lambda}\Sigma_{\lambda\mu}-i2x_{\mu}x^{\rho}\partial_{\rho}+ix^{2}\partial_{\mu})\phi(x),\nonumber \end{align}
 where the finite-dimensional matrices $\Sigma$ act in the space
of $\phi$'s Lorentz indices; they have to satisfy the commutation
relation (notice the sign difference from the first equation in (\ref{eq:CA}))\[
[\Sigma_{\mu\nu},\Sigma_{\rho\sigma}]=+i(\eta_{\mu\rho}\Sigma_{\nu\sigma}\pm\mbox{perms})\ .\]

The algebra (\ref{eq:CA}) corresponds to the mostly minus Minkowski
signature. Beware that the literature uses inconsistent sign conventions
for various generators, in particular $M_{\mu\nu}$ and $D$ (our
conventions are those of \cite{Ferrara:1973eg}). Also, the generator action is usually
given with relative sign errors among various terms; this is not surprising
because in practice these expressions are actually rarely used. However,
we will need them, so we re-checked from scratch by using the original
method of Mack and Salam \cite{Mack:1969rr}.

As is well known, the algebra (\ref{eq:CA}) is isomorphic to $SO(d,2)$.
The isomorphism is exhibited by identifying
\begin{gather*}
J_{\mu\nu}=M_{\mu\nu},\qquad J_{d,d+1}=D\ \\
J_{d,\mu}=\frac{1}{2}(P_{\mu}-K_{\mu}),\qquad J_{d+1,\mu}=\frac{1}{2}(P_{\mu}+K_{\mu})\ ,\end{gather*}
 and then $J_{\alpha\beta}$ $(\alpha,\beta=0\ldots d+1)$ satisfy
the $SO(d,2)$ commutation relations
\begin{align*}
[J_{\alpha\beta},J_{\gamma\delta}] =-i(\eta_{\alpha\gamma}J_{\beta\delta}\pm\mbox{perms})\ ,
\qquad\eta_{\alpha\beta} =\mbox{diag}(+,-,\ldots,-;-,+)\ .
\end{align*}

\section{Unitarity bound for vectors}

\label{sec:u-bound}Metsaev \cite{Metsaev95} and Minwalla \cite{Minwalla:1997ka} have shown that
in any number of spacetime dimensions $d$ a unitary primary vector
must have dimension $\Delta\geq d-1$. We will give here an independent
derivation of this result (see \cite{Grinstein:2008qk}, \cite{Dorigoni:2009ra} for similar arguments)
based on the fact that conformal invariance
fixes the primary vector two-point function to have the (Euclidean)
form:\begin{equation*}
\left\langle Y_{\mu}(x)Y_{\nu}(0)\right\rangle =\frac{1}{(x^{2})^{\Delta}}\left(\delta_{\mu\nu}-2\frac{x_{\mu}x_{\nu}}{x^{2}}
\right).\end{equation*}
 Write this as a sum of derivatives:\begin{equation*}
\left\langle Y_{\mu}(x)Y_{\nu}(0)\right\rangle =\left(1-\frac{1}{\Delta}\right)\frac{1}{(x^{2})^{\Delta}}\delta_{\mu\nu}-\frac{1}{2(\Delta-1)\Delta}\partial_{\mu}\partial_{\nu}\frac{1}{(x^{2})^{\Delta-1}}\end{equation*}
 and pass to momentum space by using:
 \begin{equation*}
\frac{1}{(x^{2})^{\Delta}}\rightarrow \mathrm{const}\frac{\Gamma(d/2-\Delta)}{4^{\Delta}\Gamma(\Delta)}(p^{2})^{\Delta-d/2},\label{eq:FT}\end{equation*}
 where $\mathrm{const}$ does not depend on $\Delta.$ Keeping track only of
the relative factor among the two tensor structures, we have:
\begin{align*}
\left\langle Y_{\mu}(p)Y_{\nu}(-p)\right\rangle  \propto\left(B_{1}\delta_{\mu\nu}+B_{2}\frac{p_{\mu}p_{\nu}}{p^{2}}
\right)(p^{2})^{\Delta-d/2},\qquad B_{1}=1,\ B_{2}=-2\frac{\Delta-d/2}{\Delta-1}\ .
\end{align*}
The spectral density of the Wightman function in the forward Minkowski cone can be extracted from the Euclidean spectral density via the Wick rotation. We have:
\begin{equation*}
\left\langle 0|Y_{\mu}(p)Y_{\nu}(-p)|0\right\rangle \propto
-\theta(p^0) \theta (p^2)
\left(B_{1}\eta_{\mu\nu}+B_{2}\frac{p_{\mu}p_{\nu}}{p^{2}}\right)
(p^{2})^{\Delta-d/2}
 \end{equation*}
(where we fixed the sign by matching the spatial components). Unitarity implies that for any complex constant four-vector $\chi$$_{\mu}$,
the contraction $\chi.Y$ must have a positive spectral density, i.e.
\begin{equation*}
-B_{1}\chi.\chi^{\dagger}-B_{2}\frac{|\chi.p|^{2}}{p^{2}}\geq0,\quad\label{eq:unit}
\end{equation*}
 for any $\chi$ and any $p$ in forward cone. Taking $\vec{p}=0$,
$p_{0}=1$ we get:
\begin{align*}
-B_{1}(|\chi_{0}|^{2}-|\vec{\chi|^{2}})-B_{2}|\chi_{0}|^{2} \geq0\quad\Longleftrightarrow\quad B_{1}\ge0,B_{1}+B_{2}\le0
\quad
\Longleftrightarrow\quad
\Delta\geq d-1.
\end{align*}
 Notice that
for $\Delta=d-1$ the spectral density of $\partial_{\mu}Y^{\mu}$
vanishes. Thus a dimension $d-1$ vector primary is a conserved current,
just like in $d=4$.

\section{Energy-momentum tensor of the $\xi$-gauge Maxwell theory}
\label{sec:EMxi}
Here we show that the energy-momentum tensor of the gauge fixed Maxwell theory (\ref{eq:Lxi}) can be improved to be traceless, for $\xi=\xi_*$. 
This gives an alternative proof of the conformal invariance at this particular value of the gauge fixing parameter.

For a general $\xi$, the energy-momentum tensor and its trace are given by 
\begin{eqnarray*}
T_{\mu\nu}=\frac{\eta_{\mu\nu}}{4}(F_{\rho\sigma})^2-F_{\mu}^{\ \rho}F_{\rho\nu}+\xi^{-1}\left(A_{\mu}\partial_{\nu}(\partial A)+A_{\nu}\partial_{\mu}(\partial A)-\eta_{\mu\nu}\left(A^{\rho}\partial_{\rho}(\partial A)+(\partial A)^{2}/2\right)\right)\ ,\\
 T^{\mu}_{\ \mu}=\left(\frac{d}{2}-2\right)\left(\partial_{\mu}A_{\nu}\partial^{\mu}A^{\nu}-\partial_{\mu}A_{\nu}\partial^{\nu}A^{\mu}\right)+\xi^{-1}\left((2-d)A_{\mu}\partial^{\nu}\partial^{\mu}A_{\nu}-\frac{d}{2}(\partial A)^{2}\right)\ .
 \end{eqnarray*}
The condition under which an improvement is possible was discussed in Section \ref{sec:conf.lost}. In the present case, the trace of
the energy-momentum tensor must take the form 
\begin{align*}
T^{\mu}_{\ \mu}=\alpha\, \partial^2(A_{\mu}A^{\mu})+\beta\,\partial_{\mu}\partial_{\nu}(A^{\mu}A^{\nu})\ ,
\end{align*}
 up to the equations of motion: 
 \begin{align*}
\partial^2A_{\mu}+\frac{1-\xi}{\xi}\partial_{\mu}(\partial A)=0\ .
\end{align*}
 By collecting coefficients of four independent terms $(\partial A)^{2}$,
$A_{\mu}\partial^{\mu}(\partial A)$, $\partial_{\nu}A_{\mu}\partial^{\nu}A^{\mu}$,
and $\partial_{\nu}A^{\mu}\partial_{\mu}A^{\nu}$, we obtain 
\begin{align*}
\beta=-\frac{d}{2\xi}\ ,\ \ \frac{d}{2}-2=2\alpha=-\beta\ ,\ \ \frac{2-d}{\xi}=2\alpha\left(1-\frac{1}{\xi}\right)+2\beta\ .
\end{align*}
 These equations have one and only solution 
$\xi=\xi_*$.

\end{document}